\documentclass[aps,pra,twocolumn,superscriptaddress,showpacs]{revtex4}
\usepackage{bm}
\usepackage{epsf}
\usepackage{amssymb}
\usepackage{amsmath}
\usepackage{graphicx}
\usepackage{rotating}
\usepackage{epsfig}
\usepackage{psfrag}
\usepackage{amsmath}
\usepackage{hyperref}

\hypersetup{
    bookmarks=true,         
    unicode=false,          
    pdftoolbar=true,        
    pdfmenubar=true,        
    pdffitwindow=true,      
    pdftitle={My title},    
    pdfauthor={Author},     
    pdfsubject={Subject},   
    pdfcreator={Creator},   
    pdfproducer={Producer}, 
    pdfkeywords={keywords}, 
    pdfnewwindow=true,      
    colorlinks=false,       
    linkcolor=red,          
    citecolor=green,        
    filecolor=magenta,      
    urlcolor=cyan           
}

\DeclareMathAlphabet{\bi}{OML}{cmm}{b}{it}

\begin{document}

\def\ba{\mathbf{a}}
\def\d{\mathbf{d}}
\def\P{\mathbf{P}}
\def\bF{\mathbf{F}}
\def\bk{\mathbf{k}}
\def\bkn{\mathbf{k}_{0}}
\def\bx{\mathbf{x}}
\def\bfn{\mathbf{f}}
\def\bg{\mathbf{g}}
\def\bj{\mathbf{j}}
\def\bR{\mathbf{R}}
\def\br{\mathbf{r}}
\def\bu{\mathbf{u}}
\def\bq{\mathbf{q}}
\def\bw{\mathbf{w}}
\def\bp{\mathbf{p}}
\def\bG{\mathbf{G}}
\def\bz{\mathbf{z}}
\def\bs{\mathbf{s}}
\def\E{\mathbf{E}}
\def\bv{\mathbf{v}}
\def\b0{\mathbf{0}}
\def\la{\langle}
\def\ra{\rangle}
\def\beq{\begin{equation}}
\def\eeq{\end{equation}}
\def\bea{\begin{eqnarray}}
\def\eea{\end{eqnarray}}
\def\bdm{\begin{displaymath}}
\def\edm{\end{displaymath}}
\def\bnab{\bm{\nabla}}
\def\bA{{\bf A}}
\def\R{R^0_{\mathrm{TF}}}
\def\L{\bm{\Lambda}}
\def\bua{{\mathbf{u}}^{\mathrm{ad}}}
\def\bus{{\mathbf{u}}^{\mathrm{en}}}

\title{First and second sound in a strongly interacting Fermi gas}
\author{E.~Taylor}\altaffiliation{Current address: Department of Physics, The Ohio State University, Columbus, Ohio, 43210}
\affiliation{CNR-INFM BEC Center and Dipartimento~di~Fisica, Universit\`a di Trento, I-38050 Povo, Trento, Italy}
\author{H.~Hu}
\affiliation{ACQAO and Centre for Atom Optics and Ultrafast Spectroscopy, Swinburne University of Technology, Melbourne, Victoria 3122, Australia}
\affiliation{Department of Physics, Renmin University of China, Beijing 100872, China}
\author{X.-J.~Liu}
\affiliation{ACQAO and Centre for Atom Optics and Ultrafast Spectroscopy, Swinburne University of Technology, Melbourne, Victoria 3122, Australia}
\author{L.~P.~Pitaevskii}
\affiliation{CNR-INFM BEC Center and Dipartimento~di~Fisica, Universit\`a di Trento, I-38050 Povo, Trento, Italy}
\affiliation{Kapitza Institute for Physical Problems, Kosygina 2, 119334 Moscow, Russia}
\author{A.~Griffin}
\affiliation{Department of Physics, University of Toronto, Toronto, Ontario,
Canada, M5S 1A7}
\author{S.~Stringari}
\affiliation{CNR-INFM BEC Center and Dipartimento~di~Fisica, Universit\`a di Trento, I-38050 Povo, Trento, Italy}

\date{November 3, 2009}

\begin{abstract}
Using a variational approach, we solve the equations of two-fluid hydrodynamics for a uniform and trapped Fermi gas at unitarity.  In the uniform case, we find that the first and second sound modes are remarkably similar to those in superfluid Helium, a consequence of strong interactions.  In the presence of harmonic trapping, first and second sound become degenerate at certain temperatures.  At these points, second sound hybridizes with first sound and is strongly coupled with density fluctuations, giving a promising way of observing second sound.  We also discuss the possibility of exciting second sound by generating local heat perturbations.    
\end{abstract}

\maketitle

\section{Introduction}
First and second sound are a spectacular manifestation of superfluidity in strongly interacting quantum liquids.  They describe the coupled oscillations of the superfluid and normal fluid components at finite temperatures. In $^4$He, second sound is a temperature oscillation, in contrast to first sound, which is a density oscillation. That second sound should exist in superfluids was first pointed out by Tisza~\cite{Tisza}.  A complete theory based on the equations of two-fluid hydrodynamics was derived by Landau~\cite{Landau41}.  In conjunction with this theory, detailed measurements of the speed of second sound were crucial for developing a microscopic understanding of superfluid $^4$He.  Two-component Fermi gases close to unitarity~\cite{TrentoFermiReview,PethickSmithBook} provide a new type of strongly interacting superfluid.  Being a dilute gas, the unitary Fermi superfluid exhibits unique ``universal" properties shared with neutron matter and other exotic systems~\cite{Gezerlis08,Schafer09}.  An understanding of first and second sound in this system will be crucial to measuring quantities of interest such as transport coefficients, the superfluid density, and the finite temperature equation of state.

Many classic signatures of superfluidity have already been seen in ultracold gases~\cite{TrentoBook,TrentoFermiReview,PethickSmithBook} including quantized vortices, the absence of viscosity, and the Josephson effect.  Second sound has not yet been observed, however.  Its detection requires that one be able to access the collisional hydrodynamic regime where the two-fluid equations are valid.  For Bose-condensed atomic gases, this is difficult because of the small value of the $s$-wave scattering length.  However, this requirement is easily satisfied in superfluid Fermi gases close to unitarity where the $s$-wave scattering length is infinite.  Due to the strong interactions characterizing the unitary regime, scattering between thermal excitations is sufficiently rapid to be in the hydrodynamic regime~\cite{Thomas05,Massignan05,Wright07}.  At the same time, in strongly interacting atomic Fermi gases, second sound is weakly coupled to density perturbations.  Modulating the frequency of the harmonic potential, for instance, will not appreciably excite second sound. Thus, even if experiments can reach the hydrodynamic regime, the ability to excite and detect second sound in trapped gases remains a significant challenge.  

In this article, we solve the Landau two-fluid equations of uniform and trapped Fermi superfluids at unitarity, presenting both numerical and analytical solutions.  The nonuniformity of the equilibrium thermodynamic functions in a trapped gas makes a reliable solution of these equations very challenging.  First solutions were obtained Ref.~\cite{He07} using brute force methods with simplified thermodynamic functions and in Ref.~\cite{TaylorPRA08} using a very simple ansatz for the velocity fields $\bv_s(\br,t)$ and $\bv_n(\br,t)$.  These papers did not, however, capture the basic features of the problem which represent the major achievement of the present paper: the surprising analogy between the unitary Fermi gas and superfluid Helium as a result of strong interactions, the strong dependence of second sound on the behaviour of the thermodynamic functions both at low $T$ and close to $T_c$, and the peculiar bimodal structure in the density response arising from the hybridization between first and second sound in a harmonic trap.  In this work, we develop the variational approach to solve the two-fluid equations and illustrate these features.  We also discuss the possibility of exciting second sound by generating local heat perturbations.

\section{First and second sound} 
The dissipationless Landau two-fluid equations in a trap $V_{\mathrm{ext}}$ are given by~\cite{Landau41,ZNGBook,TaylorPRA05}:
\bea \frac{\partial \rho}{\partial t} + \bnab\cdot\bj = 0,\;\frac{\partial s}{\partial t} + \bnab\cdot(s \bv_n) = 0, \label{cont}\eea
\bea m\frac{\partial\bv_s}{\partial t} 
=-\bnab\left(\mu + V_{\mathrm{ext}}\right), \;
\frac{\partial \bj}{\partial t} &=& -\bnab P - n\bnab V_{\text{ext}}. \label{jt}\eea
Here, $\bj = \rho_s\bv_s+\rho_n\bv_n$ is the current density, $\rho_s$ and $\rho_n$ are the superfluid and normal fluid densities for a gas with total mass density $\rho \equiv mn = \rho_s+\rho_n$.  $P = -U  + Ts + \mu n$ is the local pressure of the gas with energy density $U$ and entropy density $s$, while $\mu(n)$ is the local chemical potential.  Since we are interested only in the linear solutions of these hydrodynamic equations, we have omitted terms which are quadratic in velocity in (\ref{jt}).  

One can show~\cite{TaylorPRA05,TaylorPRA08} that the normal mode solutions of the hydrodynamic equations with frequency $\omega$ can be derived  by minimizing the variational expression
\bea \lefteqn{\omega^2 =}&&\nonumber\\&&\! \frac{\int \!\!d\br\left[\!\frac{1}{\rho_0}\!\left(\frac{\partial
P}{\partial\rho}\right)_{\!\bar{s}}\!(\delta\rho)^2\! +\! 2\rho_0\!\left(\frac{\partial
T}{\partial\rho}\right)_{\!\bar{s}}\!\delta\rho\delta \bar{s}\!+\! \rho_0\left(\frac{\partial
T}{\partial \bar{s}}\right)_{\!\rho}\!(\delta \bar{s})^2\right]}{\int d\br\;\left[\rho_{s0}\bu^2_s(\br) + \rho_{n0}\bu^2_n(\br)\right]}\nonumber\\ \label{TFVar}\eea
with respect to the displacement fields ${\bu}_n$ and ${\bu}_s$ characterizing the density and entropy fluctuations according to $\delta\rho = -\bnab\cdot[\rho_{s0}\bu_s + \rho_{n0}\bu_n]$ and $\delta \bar{s} = -\bu_n\cdot\bnab\bar{s}_0+(\bar{s}_0/\rho_0)\bnab\cdot[\rho_{s0}(\bu_s-\bu_n)]$.  
These are related to the normal and superfluid velocity fields by  $\dot{\bu}_s = \bv_s$, $\dot{\bu}_n = \bv_n$.  $\bar{s}\equiv s/\rho$ is the entropy per unit mass.  The effect of the trapping potential enters (\ref{TFVar}) through the position dependent equilibrium thermodynamic functions.  These functions are calculated using the theory developed in Ref.~\cite{HLD}, which is an improved version~\cite{note4} of the well known theory of Nozi\`eres and Schmitt-Rink (NSR)~\cite{NSR}.  We hope our work encourages more \textit{ab initio} Monte-Carlo calculations of thermodynamic functions~\cite{Bulgac07} in order to improve the accuracy of the theoretical predictions for the propagation of second sound.

Before discussing the general solutions of (\ref{TFVar}), it is useful to solve this equation using ansatzes for 
a pure in-phase mode [$\bu_s(\br,t)=\bu_n(\br,t)\equiv \bu^{(1)}(\br,t)$] and a pure out-of-phase mode [$\rho_{s0}(\br)\bu^{(2)}_s(\br,t)=-\rho_{n0}\bu^{(2)}_n(\br,t)$], hereafter referred to as first and second sound.  One can prove that these two modes satisfy an exact orthogonality condition obeyed by the hydrodynamic equations (see Appendix~\ref{orthosec}).
These first and second sound modes correspond to pure density ($\delta T(\br,t) = 0$) and temperature ($\delta\rho(\br,t)=0$) oscillations, respectively~\cite{TaylorPRA08}.  They are the exact variational solutions of (\ref{TFVar}) when the coefficient $(\partial T/\partial \rho)_{\bar{s}}$ in (\ref{TFVar}) is set to zero.  A nonzero value for this quantity leads to coupling between density and temperature oscillations and hence, between first and second sound.  
A convenient way of describing this coupling is in terms of the dimensionless Landau--Placzek (LP) ratio
$\epsilon \equiv \frac{\bar{c}_p-\bar{c}_v}{\bar{c}_v}$, where $\bar c_p=T(\partial\bar s/\partial T)_P$
and $\bar{c}_v = T(\partial\bar{s}/\partial T)_{\rho}$ are the equilibrium specific heats per unit mass at constant pressure and density, respectively.   When $(\partial T/\partial\rho)_{\bar{s}}=0$, this means $\bar{c}_p = \bar{c}_v$, or $\epsilon=0$.  We find that--similarly to liquid $^4$He--the solutions of the two-fluid equations for a trapped Fermi gas are well described by weakly coupled first and second sound modes, even though $\epsilon$ can be order unity in Fermi gases.

Inserting the ansatz $\bu_s(\br,t) = \bu_n(\br,t)\equiv \bu^{(1)}(\br,t)$ for first sound into (\ref{TFVar}), taking the variation with respect to $\bu^{(1)}$ and making use of standard thermodynamic identities as well as the equilibrium conditions $\bnab P_0 = -n_0\bnab V_{\mathrm{ext}}$ and $\bnab T_0 = (\partial T/\partial\rho)_{\bar{s}}\bnab\rho_0 + (\partial T/\partial\bar{s})_{\rho}\bnab\bar{s}_0=0$, one obtains Euler's equation
\bea m\omega^2_1\bu^{(1)}\!\! &=&\! -\frac{m}{\rho_0}\bnab\left[\!\rho_0\left(\frac{\partial P}{\partial \rho}\right)_{\!\bar{s}}\!\!\bnab\cdot\bu^{(1)}\right]\!+\!\bnab(\bu^{(1)}\!\cdot\!\bnab V_{\mathrm{ext}})\nonumber\\&&- (\bnab\cdot\bu^{(1)})\bnab V_{\mathrm{ext}}.\label{Eulerv}\eea   
For a Fermi gas at unitarity, where $P=2U/3$, it follows that $\rho_0(\partial P/\partial \rho)_{\bar{s}} = 5P_0/3$, a result also satisfied by a noninteracting gas.  Using this in (\ref{Eulerv}), one sees that it is identical to the equation for a collisionally hydrodynamic Bose gas above $T_c$ using ideal gas thermodynamics~\cite{Griffin97,ZNGBook}.

For second sound, we use the ansatz $\bu^{(2)}_s(\br,t)\rho_{s0}(\br)=-\bu^{(2)}_n(\br,t)\rho_{n0}(\br)$.  
Inserting this into (\ref{TFVar}), the variational procedure gives the following equation for the superfluid displacement field:
\bea \omega^2_2\bu^{(2)}_s = -\frac{s_0}{\rho_0}\bnab\left[\frac{1}{\rho_0}\left(\frac{\partial T}{\partial \bar{s}}\right)_{\!\rho}\bnab\cdot\left(\frac{s_0\rho_{s0}}{\rho_{n0}}\bu^{(2)}_s\right)\right].\label{Entropyv}\eea
Since $\delta\rho(\br,T)=0$ for this mode, one may also write this as a closed equation for the temperature fluctuations $\delta T(\br) = (\partial T/\partial \bar{s})_{\rho}\delta \bar{s}(\br)$, the entropy fluctuations $\delta \bar{s}$ being related to $\bu^{(2)}_s$ by the expression below (\ref{TFVar}) and the second sound ansatz.

\section{First and second sound in a uniform Fermi gas superfluid} 
\label{uniformsec}

\begin{figure}
\epsfig{file=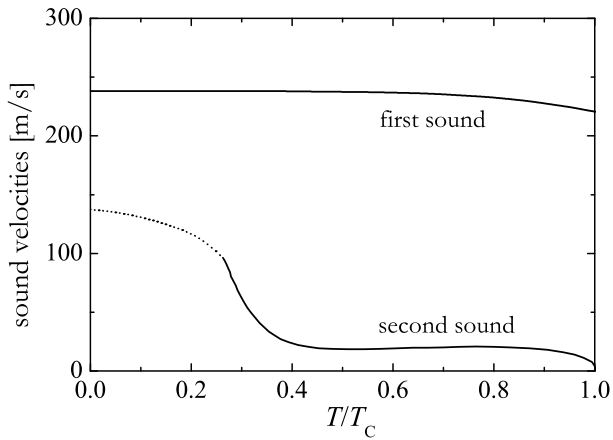, angle=0, width=0.45\textwidth}
\caption{Experimental values of first and second sound speeds in superfluid $^4$He. Calculating the 
thermodynamic functions in (\ref{soundspeeds}) using the phonon-roton excitation 
spectrum measured by neutron scattering, one gets excellent agreement with 
this data (from Refs.~\cite{Atkins}and \cite{Peshkov60}).}
\label{Hesoundfig}
\end{figure}

For a uniform superfluid ($V_{\mathrm{ext}}=0$), the solutions of (\ref{Eulerv}) and (\ref{Entropyv}) are plane waves of  wavevector $q$ with dispersion $\omega_1=c_1q$ and $\omega_2=c_2 q$, where
\bea c^2_1 = \left(\frac{\partial P}{\partial\rho}\right)_{\!\bar{s}},\; c^2_2 = T\frac{\bar{s}^2_0}{\bar{c}_v}\frac{\rho_{s0}}{\rho_{n0}}.\label{soundspeeds}\eea
These first and second sound velocities are the standard expressions~\cite{Landau41,ZNGBook} used to describe superfluid $^4$He where $\epsilon \ll 1$ except in a narrow region around $T_c$. The measured sound velocities, shown in Fig.~\ref{Hesoundfig} for superfluid $^4$He, as a function of temperature~\cite{Atkins,Peshkov60}, agree with (\ref{soundspeeds}).

\begin{figure}
\epsfig{file=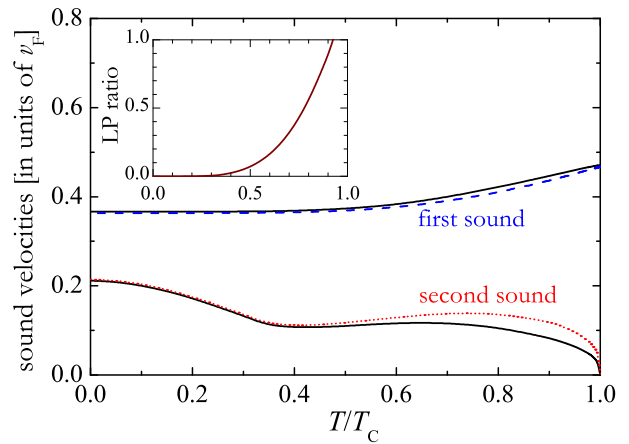, angle=0, width=0.45\textwidth}
\caption{Two-fluid sound speeds in a uniform Fermi gas at unitarity (solid lines).  The uncoupled ($\epsilon=0$) first and second sound speeds given by  (\ref{soundspeeds}) are shown as dashed and dotted lines, respectively.  Inset: Landau--Placzek ratio $\epsilon$ as a function of $T/T_c$.}
\label{soundfig}
\end{figure}

In Fig.~\ref{soundfig}, we plot the calculated values of $c_1$ and $c_2$ (shown by dashed lines) given by (\ref{soundspeeds}) for a uniform Fermi superfluid gas at unitarity, using NSR thermodynamics~\cite{HLD,note4,NSR}.  In this figure, we also show the full solutions of the two-fluid equations, (\ref{TFVar}).  One immediately notes that the full solutions are well described by the uncoupled first and second sound modes, (\ref{soundspeeds}).   To understand the effect of coupling (due to a finite value of $(\partial T/\partial\rho)_{\bar{s}}$ in (\ref{TFVar})), we write the solution of the full two-fluid equations (\ref{TFVar}) as a linear combination of first ($\bu^{(1)}$) and second sound ($\bu^{(2)}_{s,n}$) modes.  The resulting variational solutions of (\ref{TFVar}) in a uniform superfluid are sound modes with dispersion~\cite{NozieresPines2} 
\bea \tilde{\omega}^2_{1,2} = \frac{\omega^2_1+\omega^2_2 \pm \sqrt{(\omega^2_1-\omega^2_2)^2 + \delta^4}}{2},\label{solntf}\eea where $\delta^4 = 4q^4\bar{s}^2_0(\rho_{s0}/\rho_{n0})(\partial P/\partial s)^2_{\rho} = 4\omega^2_1\omega^2_2\epsilon/(1+\epsilon)$ determines the coupling strength.  (Note that $(\partial P/\partial s)_{\rho} = \rho_0(\partial T/\partial\rho)_{\bar{s}}$.)  This shows that the coupling between first and second sound will be small as long as $(c^2_1-c^2_2)^2/4c^2_1c^2_2\gg \epsilon/(1+\epsilon)$, a condition met even if $\epsilon$ is not small due to the fact that the speeds of first and second sound are never very close~\cite{note}.

While the speed $c_1$ of first sound shown in Fig.~\ref{soundfig} is weakly dependent on temperature, the speed $c_2$ of second sound varies significantly between $T=0$ and $T_c$, where it vanishes (since $\rho_s=0$).  
Comparing Figs.~\ref{Hesoundfig} and \ref{soundfig}, the similarity in the temperature dependence of $c_2(T)$ for a superfluid $^4$He and a uniform Fermi superfluid gas at unitarity is quite striking.  In both systems, Goldstone phonons are the dominant thermal excitations at low temperatures over a wide range of temperatures, which leads to the sharp increase in $c_2(T)$ as $T\to 0$.  Using (\ref{soundspeeds}) and the leading contributions to thermodynamic functions due to thermally excited phonons, we recover the celebrated results~\cite{Landau41} $c_1 = c$ and $c_2=c/\sqrt{3}$, where in our case $c=\sqrt {\xi/3}v_F \sim 0.37 v_F$ is the $T=0$ value of the speed of sound~\cite{TrentoFermiReview}.  Our NSR theory recovers this $T=0$ result.   If one instead used thermodynamics based on BCS Fermi excitations alone (without Goldstone phonons), one obtains the incorrect result that $c_2$ vanishes as $T\to 0$.  The fact that phonons dominate thermodynamics up to such high temperatures ($T\sim 0.4T_c$) is unusual for a superfluid \textit{gas} and results from the strong interactions exhibited by the unitary Fermi gas~\cite{Combescot06}.  These results only give the limiting low-$T$ behavior of the two-fluid equations and are not valid at very low $T$ where hydrodynamics breaks down.  At higher temperatures ($T\gtrsim 0.5T_c$) thermodynamics is dominated by high energy rotons in superfluid $^4$He and by BCS quasiparticles (which have a large energy gap) in the unitary Fermi gas.   The two-fluid modes discussed here are very different from those in dilute Bose-condensed gases where the weak interactions give rise to a strong coupling between density and entropy oscillations at all but the lowest temperatures~\cite{ZNGBook,TrentoBook}.

\section{First and second sound in a trapped Fermi gas superfluid} 

\begin{figure}
\epsfig{file=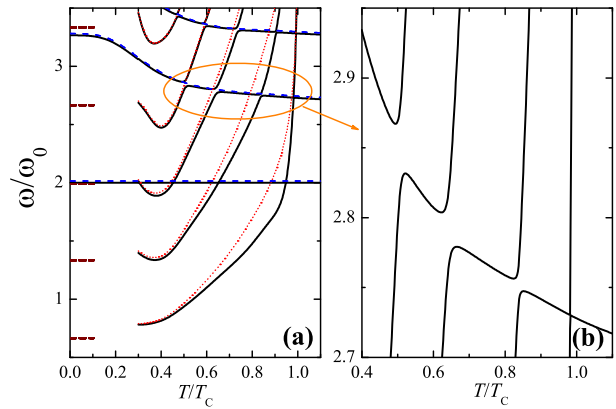, angle=0, width=0.45\textwidth}
\caption{Left panel: Two-fluid modes of a unitary Fermi gas in an isotropic harmonic trap of frequency $\omega_0$.  The full solutions of the two-fluid equations are given by the solid lines. The lowest first sound modes (blue dashed lines) given by (\ref{Eulerv}) are shown along with the lowest second sound modes (red dotted lines) given by (\ref{Entropyv}).    The first sound mode frequencies $\omega_1(n)$ are indistinguishable from the full solutions away from the hybridization points and have been shifted upwards slightly for clarity.  The dashed lines at $T=0$ show our analytic results for the second sound frequency $\omega_2(n)$ discussed in the text.  Right panel:  A blow-up of part of the left panel showing the hybridization between second sound and the $n=1$ first sound mode.}
\label{dispersionfig}
\end{figure}

We now consider the case of a unitary Fermi gas confined by a harmonic trapping potential.  For clarity, we only consider the breathing modes ($l=0$) for an isotropic potential $V_{\mathrm{ext}} = \frac{1}{2}m\omega^2_0r^2$.  The extension to other types of modes (dipole, quadrupole, etc.) and to anisotropic potentials is straightforward in our approach.  In Fig.~\ref{dispersionfig}, we show our results for the breathing mode frequencies of the full two-fluid equations using a variational polynomial ansatz (see Appendix~\ref{ansatzsec}) for $\bu_s(\br)$ and $\bu_n(\br)$ in (\ref{TFVar}).  Also shown are the frequencies of the lowest first [$\omega_1(n)$] and second sound [$\omega_2(n)$] modes, given by (\ref{Eulerv}) and (\ref{Entropyv}) ($n=0,1,...$ is the number of radial nodes).  First sound solutions are the ``horizontal" branches while the dotted lines are the second sound solutions.  Figure~\ref{dispersionfig}(a) shows that the frequencies of the uncoupled first and second sound modes are close to the full solutions of (\ref{TFVar}), showing that, similar to the uniform case, the two-fluid modes of a \textit{trapped} unitary Fermi gas are almost pure density and temperature oscillations.

The two-fluid equations in (\ref{cont}) and (\ref{jt}) simplify to those of a pure superfluid ($\rho_{n0}=0$) at $T=0$ and to a normal fluid ($\rho_{s0}=0$) above $T_c$.  In both cases, there is a single fluid, whose displacement field is given by (\ref{Eulerv}).  In these two limits, (\ref{Eulerv}) gives analytic solutions for the first sound breathing mode frequencies.  At $T=0$, the hydrodynamic equation of motion for a trapped unitary Fermi superfluid~\cite{Stringari04EPL} predicts breathing modes with frequencies~\cite{Amoruso99,Bruun99} $\omega_1(n)= \omega_0\sqrt{4(n+1)(n+3)/3}$.   
Above $T_c$, our numerical results shown in Fig.~\ref{dispersionfig}(a) for the first sound breathing mode frequencies quickly approach the analytic result for a collisionally hydrodynamic classical gas~\cite{Bruun99}, $\omega_1(n) =\omega_0\sqrt{10n/3+4}$.  For the $n=0$ first sound mode, both expressions reduce to the same value $\omega_1(0)=2\omega_0$.  The $n=0$ first sound scaling solution $\bu^{(1)}\propto \br$ is actually an exact solution of the full two-fluid hydrodynamic equations at unitarity with this frequency at all temperatures~\cite{He07,TaylorPRA08} (moreover, it is an \textit{exact} solution of the many-body Schr\"odinger equation at unitarity~\cite{Castin04} for an isotropic trap).   In contrast, the $n=1$ and higher first sound modes vary strongly with temperature.  The measurement of the frequencies of these modes would be of great interest since they are sensitive to the temperature dependence of the equation of state of the Fermi gas.

One can also derive a simple analytic results for the second sound frequencies $\omega_2(n)$ as $T\to 0$ and also $T\to T_c$.  For the low temperature region, our approach is analogous to that used to obtain the uniform gas result $c_2 = c/\sqrt{3}$. Assuming that phonons dominate thermodynamics, we solve (\ref{Entropyv}) analytically within LDA.  One finds (see Appendix~\ref{entropymodesec}) $\omega_2(n)=\frac{2}{3}(n+1)\omega_0$.  
The numerical calculations of $\omega_2(n)$ in Fig.~\ref{dispersionfig}(a) are difficult for  $T\lesssim 0.4T_c$ due to the smallness of the normal fluid density.  However, we note that our low $T$ results ($T\lesssim 0.4T_c$) extrapolate to our analytic results at $T= 0$.  As noted in Sec.~\ref{uniformsec}, when $T\to 0$, the predictions of the two-fluid hydrodynamic equations are only useful in giving the limiting low-$T$ behavior, the hydrodynamic regime for the thermal component no longer being ensured.  

As $T\to T_c$, the behaviour of the second sound frequencies is very peculiar.  Recall that the velocity of second sound in a uniform superfluid vanishes at $T_c$ (see Fig.~\ref{soundfig}) because $\rho_{s0}\to 0$.  The behaviour of second sound in a trap, on the other hand, depends sensitively on the spatial dependence of $\rho_{s0}(\br)$ close to $T_c$.  Within LDA, the size $R_s = R_{TF}\sqrt{1-T/T_c}$ of the superfluid component vanishes as $T\to T_c$, giving rise to an increase in the minimum wavevector, $q\sim 1/R_s$.  Using $\rho_{s0} \propto [1-T/T_c]^{2\beta}$, we find that $\omega_2 \sim c_2 q \propto [1-T/T_c]^{\beta-1/2}$.  For $\beta < 1/2$, $\omega_2$ diverges.  The $\rho_s$ data we use~\cite{note4} are consistent with the correct value $\beta \simeq 1/3$~\cite{Josephson} and consequently, we expect (and find) our calculated LDA values of $\omega_2$ diverge as $T\to T_c$.  Our results emphasize that the low ($T\simeq 0$) and high ($T\simeq T_c$) temperature behaviour of the second sound mode frequencies depend sensitively on the thermodynamic properties of the unitary Fermi gas.

The rapid increase in the frequency of second sound as a function of temperature has the important consequence that, in contrast to the situation in a uniform Fermi gas (see Fig.~\ref{soundfig}), the first and second sound mode frequencies cross each other.  As anticipated in Ref.~\cite{He07}, this leads to a hybridization effect between first and second sound.  Our results for the solution of the \textit{full} two-fluid equations (\ref{TFVar}) [which includes the coupling between first and second sound left out of (\ref{Eulerv}) and (\ref{Entropyv})] clearly reveal the hybridization between the $n=1$ and higher first sound modes and the second sound modes (being an exact solution of the full two-fluid hydrodynamic equations at all temperatures, the $n=0$ first sound mode does not couple to second sound).  The hybridization between the $n=1$ first sound and the lowest four second sound modes is shown in Fig.~\ref{dispersionfig}(b).  These results are easily understood by looking for variational solutions of the full two-fluid equations (\ref{TFVar}) as a superposition $\tilde{\bu}_{1,2} = A\bu^{(1)}+B\bu^{(2)}_{s,n}$ of first and second sound.  Varying (\ref{TFVar}) with respect to $A$ and $B$, we obtain again the dispersion (\ref{solntf}), where $\omega_1$ and $\omega_2$ are the solutions of (\ref{Eulerv}) and (\ref{Entropyv}), while the coupling strength is now
\bea \delta^2 \equiv\frac{\int_{R_s} d\br\;(\partial P/\partial s)_{\rho}\bnab\cdot(s_0\bu^{(2)}_n)\bnab\cdot\bu^{(1)}}{\sqrt{M(1)M(2)}}.\label{coupling}\eea
Here, the integration is restricted to the superfluid region $r<R_s$ and we define $M(i)\!\equiv\!\frac{1}{2}\!\int\! d\br[\rho_{s0}(\bu^{(i)}_s)^2\!+\rho_{n0}(\bu^{(i)}_n)^2]$. The splitting between the two modes at hybridization, where $\omega_1=\omega_2$, is given by $\tilde{\omega}^2_2-\tilde{\omega}^2_1=\delta^2$. As $T\to T_c$, this splitting decreases (see Fig.~\ref{dispersionfig}(b)) since $\delta$ vanishes at $T_c$ where $R_s\to 0$.

\section{Exciting and detecting second sound in a trapped gas}
The results in Fig.~\ref{dispersionfig}(b) show that the coupling between second sound 
with first sound is strong in a Fermi superfluid when their frequencies 
are equal. Hybridization means that second sound will couple to 
the density response at these crossing points. Adding a density perturbation of the form $\delta V(\br,t)= \lambda f(\br)e^{-i\omega t}$ to the two-fluid equations, making the same ansatz as before for $\tilde{\bu}_{1,2}$ (where now $A$ and $B$ depend on time), it is straightforward to derive the response function~\cite{NozieresPines2,ZNGBook} for this perturbation, defined by 
$ \int d\br f(\br)\delta\rho(\br,t) \equiv \lambda\chi(\omega)e^{-i\omega t}$, with imaginary part
\bea -\frac{\mathrm{Im}\chi(\omega)}{2\pi} = I^2\left[
Z_1\delta(\omega^2-\tilde{\omega}^2_1) + Z_2\delta(\omega^2-\tilde{\omega}^2_2)\right]. \label{response}\eea 
Here, $I \equiv \int d\br \rho_0\bu^{(1)}\cdot\bnab f(\br)/\sqrt{M(1)}$ characterizes the strength of the density response to a perturbation with spatial dependence $f(\br)$.  Equation (\ref{response}) is valid close to the hybridization points where the frequency of the first sound mode of interest crosses with the frequency of a second sound mode and we can ignore contributions from other modes.  The response is characterized by the two poles $\tilde{\omega}_{1,2}$, with
strengths $Z_1=\frac{\tilde{\omega}^2_1 - \omega^2_2}{\tilde{\omega}^2_1 - \tilde{\omega}^2_2}$ and 
$Z_2= 1-Z_1$. At hybridization, 
where $\omega_1=\omega_2$, we have $\tilde{\omega}^2_{1,2} = \omega^2_1 \pm \delta^2/2$ and one finds $Z_1=Z_2=1/2$. This result is extremely important from an experimental standpoint. It shows, for instance, that switching on an appropriate static  perturbation $f(\br)$ will result in a simultaneous excitation of the two modes, with the occurrence of a typical beating effect.  The response is maximized by choosing $f(\br)$ such that $\bnab f(\br) = \bu^{(1)}(\br)$.  

\begin{figure}
\epsfig{file=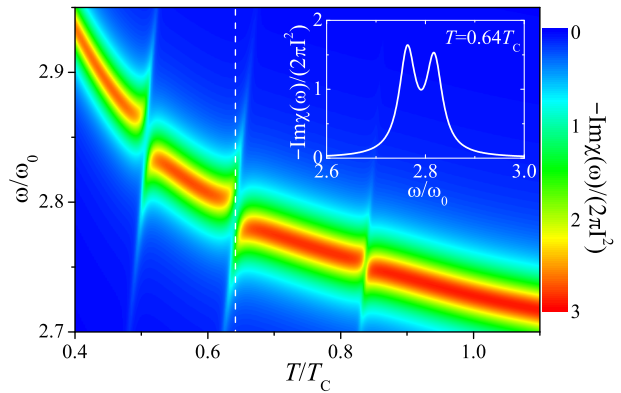, angle=0, width=0.45\textwidth}
\caption{Normalized density response $\mathrm{Im}\chi(\omega)$ of a trapped Fermi gas at unitarity to a perturbation $\delta V(\br,t)= \lambda f(\br)e^{-i\omega t}$ in the vicinity of the $n=1$ first sound mode (see text).  To simulate the effects of damping~\cite{Turlapov08}, the delta function peaks have been broadened by $0.02\omega_0$. Inset: double peak structure at the hybridization point $T\simeq 0.64T_c$ showing that first and second sound contribute equally ($Z_1\sim Z_2$) to the response.  The frequency splitting $\Delta\omega \sim 0.06\omega_0$ between peaks is large enough to be observed in experiments.}
\label{responsefig}
\end{figure}

Figure~\ref{responsefig} shows the two fluid response, clearly showing the bimodal structure in the vicinity of hybridization between the $n=1$ first sound mode and second sound. For the $n=1$ first sound mode, $f(\br)$ is well approximated by $f(\br)\simeq r^2 - \frac{1}{2}\alpha(T)r^4$ where $\alpha \equiv \langle r^2\rangle/\langle r^4\rangle$ is weakly temperature dependent.  This potential can be straightforwardly realized by modulating an optical dipole trap.  Close to hybridization, this potential excites both first and second sound, as shown by Fig.~\ref{responsefig}.  Away from hybridization, it excites only the $n=1$ first sound mode.  As discussed earlier, the measurement of this mode, characterized by the density fluctuation $\delta \rho \sim -\bnab \cdot[\rho_0\bnab f]$, would be of great interest since it is sensitive to the equation of state.   

Away from hybridization, the viscous damping of first sound should be small~\cite{Turlapov08}.  Close to hybridization, on the other hand, we expect that the damping will be enhanced.  The bulk viscosities $\zeta_1$ and $\zeta_2$ exactly vanish at unitarity~\cite{Son07} and the damping is determined by the shear viscosity $\eta$, the bulk viscosity $\zeta_3$, and the thermal conductivity $\kappa$. When $\bv_s=\bv_n$, the bulk viscosity $\zeta_3$ and the thermal conductivity $\kappa$ (recall that $\delta T=0$ for this type of motion) do not contribute (see Sec.~140 in Ref.~\cite{LLFM} and Ch.~19 in Ref.~\cite{ZNGBook} for discussions of hydrodynamic damping).  Thus, away from hybridization, only the small shear viscosity $\eta$ contributes to hydrodynamic damping of first sound.  In contrast, close to hybridization, where $\bv_s\neq \bv_n$, there will be enhanced damping arising from finite values of $\kappa$ and $\zeta_3$.  If the damping is large enough, the double peak structure in Fig.~\ref{responsefig} will be masked. However, the increased damping itself would be evidence of the excitation of second sound.  

A more direct way to excite second sound could be realized by generating a heat perturbation.  For example, a laser tuned close to an optical resonance will heat the gas.  The use of two intensity-modulated counter-propagating beams will produce a periodic heat perturbation, resulting in an additional position and time dependent term in (\ref{cont}) for the entropy density, leading to the excitation of second sound and generation of a large temperature fluctuations when the frequency and the wavelength of the perturbation are in resonance with the second sound dispersion.  This procedure would be particularly efficient in an elongated trap where one could probe the dispersion relation (\ref{soundspeeds}) of a uniform superfluid~\cite{Nikuni98}.  

\section{Conclusions}

In this paper, we have presented reliable predictions for first and second sound in a Fermi gas at unitarity, including both numerical and analytic results.  In contrast to the situation in weakly interacting Bose gases where the solutions of the two-fluid equations are, in general, a complex admixture of density and temperature oscillations (see, for instance, Refs.~\cite{ZNGBook,ZNGJLTP}), the two-fluid modes in a strongly interacting Fermi gas are weakly coupled density and temperature oscillations over a large range of temperatures.  Nonetheless, we find that in the presence of trapping, the coupling can be significant in places where hybridization occurs between first and second sound.  This effect  provides a promising way to excite and detect second sound in trapped gases by measuring the response of the system to a density probe.  Alternately, in highly anisotropic cigar-shaped traps, second sound waves could be excited directly by heating the gas with a space- and time-dependent perturbation or by a density perturbation~\cite{Arahato09}.

\section{Acknowledgments}
H.H. and X.-J.L. were supported by ARC and NSFC.  A.G. was supported by NSERC and CIFAR.  S.S. acknowledges the support of the EuroQUAM FerMix program.

\appendix

\section{Orthogonality condition}
\label{orthosec}
Suppose that $\{\bu^{(j)}_s,\bu^{(j)}_n\}$ and $\{\bu^{(k)}_s,\bu^{(k)}_n\}$ are two solutions of (\ref{TFVar}).    Inserting the first solution into the variational equations that result from (\ref{TFVar}), left-multiplying by the second solution and integrating by parts, we find
\bea \lefteqn{\omega^2_j\int d\br\;\rho_{s0}\bu^{(k)}_s\cdot\bu^{(j)}_s + \omega^2_j\int d\br\;\rho_{n0}\bu^{(k)}_n\cdot\bu^{(j)}_n=}&&\nonumber\\&&\int d\br\;\Big[\frac{1}{\rho_0}\left(\frac{\partial
P}{\partial\rho}\right)_{\!\bar{s}}\delta\rho^{(k)}\delta\rho^{(j)} + \rho_0\left(\frac{\partial
T}{\partial\rho}\right)_{\!\bar{s}}\delta\rho^{(k)}\delta \bar{s}^{(j)} \nonumber\\&&+ \rho_0\left(\frac{\partial
T}{\partial\rho}\right)_{\!\bar{s}}\delta\rho^{(j)}\delta \bar{s}^{(k)} + \rho_0\left(\frac{\partial
T}{\partial \bar{s}}\right)_{\!\rho}\delta \bar{s}^{(j)}\delta \bar{s}^{(k)}\Big].\label{bobby1}\eea
Here, $\omega_j$ is the frequency of the $\{\bu^{(j)}_s,\bu^{(j)}_n\}$ solution and $\delta\rho^{(j)},\delta s^{(j)}$ are the corresponding density and entropy fluctuations.  

We see that the right-hand side of Eq.~(\ref{bobby1}) is symmetric with respect to the exchange of the $j$ and $k$ solutions.  Thus, we obtain the orthogonality relation
\bea \left[\!\int\!\! d\br \rho_{s0}\bu^{(k)}_s\!\!\cdot\!\bu^{(j)}_s \!+\! \int d\br \rho_{n0}\bu^{(k)}_n\!\!\cdot\!\bu^{(j)}_n\!\right]\!\!\left(\omega^2_j\!-\!\omega^2_k\right)=0. \label{orthogonality}\eea
One immediately sees that as long as $\omega_1\neq \omega_2$, the existence of a first sound solution ($\bu^{(1)}_s=\bu^{(1)}_n$) implies an orthogonal  second sound solution with $\bu^{(2)}_s\rho_{s0} + \bu^{(2)}_n\rho_{n0}=0$, and vice versa.

\section{Polynomial ansatz for the displacement fields}
\label{ansatzsec}
Restricting ourselves to the breathing modes ($l=0$) in an isotropic trap, we
use the following variational ansatz for the displacement fields: 
\bea
{\bf u}_{s(n)}\left( {\bf r},t\right) &=& \br
\sum_{j=0}^{N-1}a_{s(n), j} r^{j} e^{-i\omega t},
\label{ansatz}\eea
where $N$ is the number of the variational parameters \{$a_{s,j}$, $a_{n,j}$%
\}.  Inserting this ansatz into (\ref{TFVar}), the mode frequencies are obtained by minimizing the resulting expression with respect to these $N$ parameters.  Fig.~\ref{dispersionfig} is based on $N=8$.  In actual calculations, we have found that it is better to work in terms of the variables $\bu_s-\bu_n$ and $(\rho_{s0}\bu_s+\rho_{n0}\bu_n)/\rho_0$.

\section{Low temperature limit of the second sound frequencies in a trapped Fermi superfluid}
\label{entropymodesec}
By combining the LDA formula $T_F(\br)=T_F(0)[1-(r/R_{TF})^2]$ for the Fermi temperature with standard expressions for phonon thermodynamics~\cite{LLFM}, 
we can reduce (\ref{Entropyv}) to an equation for temperature oscillations $\delta T(\br,t)$:
\bea \bar{\omega}_2^2\delta T + (1-\bar{r}^2)\nabla^2_{\bar{r}}\delta T + \bar{r}\partial_{\bar{r}}\delta T =0\label{Tfluct}.\eea
Here, we have used the dimensionless variables $\bar{r}\equiv r/R_{TF}$ and $\bar{\omega}_2\equiv 3\omega_2/\omega_0$.  The result (\ref{Tfluct}), which is valid in the low $T$ two-fluid region, is similar in structure to the equation for the density oscillations in a Bose condensate at $T=0$ based on the Gross--Pitaevskii equation~\cite{Stringari96}.  It can be solved exactly using analogous techniques.  One finds that $\delta T(\br,\omega) = \sum_{k=0}^{n+1}a_k\bar{r}^{2k}$ are polynomial solutions with the eigenvalues $\bar{\omega}_2(n) = 2(n+1)$.  Thus, in the limit of low $T$, the breathing mode second sound temperature oscillations have the simple dispersion relation $\omega_2(n)=\frac{2}{3}(n+1)\omega_0$.

\end{document}